\documentclass[aps,prl,twocolumn,superscriptaddress]{revtex4-2}
\usepackage{amsmath,latexsym,amsfonts,amssymb}
\usepackage[colorlinks=true,linkcolor=blue,citecolor=blue,urlcolor=black]{hyperref}
\usepackage{natbib}
\usepackage{graphicx}
\usepackage[utf8]{inputenc}
\usepackage{color}
\usepackage{setspace}
\usepackage{multirow}
\usepackage{array}
\usepackage{tabularx}
\usepackage{bm}
\usepackage{gensymb}
\usepackage{float}
\usepackage[left]{lineno}
\usepackage{xcolor}
\usepackage{placeins}

\newcommand{\ndzr}{Nd$_2$Zr$_2$O$_7$}
\newcommand{\ndzrx}{Nd$_2$(Zr$_{1-x}$Ti$_x$)$_2$O$_7$}
\newcommand{\prhf}{Pr$_2$Hf$_2$O$_7$}
\newcommand{\nd}{Nd$^{3+}$}
\newcommand{\TN}{$T_{\rm N}$}

\newcolumntype{Y}{>{\centering\arraybackslash}X}
\newcolumntype{C}{>{\centering\arraybackslash}c}
\newcolumntype{P}{>{\centering\arraybackslash}p}

\begin{document}

\author{M. L\'eger}
\affiliation{Institut N\'eel, CNRS and Universit\'e Grenoble Alpes, 38000 Grenoble, France}
\affiliation{Laboratoire L\'eon Brillouin, Universit\'e Paris-Saclay, CNRS, CEA, CE-Saclay, F-91191 Gif-sur-Yvette, France}
\author{E. Lhotel}
\email[]{elsa.lhotel@neel.cnrs.fr}
\affiliation{Institut N\'eel, CNRS and Universit\'e Grenoble Alpes, 38000 Grenoble, France}
\author{M. Ciomaga Hatnean}
\affiliation{Department of Physics, University of Warwick, Coventry, CV4 7AL, United Kingdom}
\author{J. Ollivier}
\affiliation{Institut Laue Langevin, F-38042 Grenoble, France}
\author{A. R. Wildes}
\affiliation{Institut Laue Langevin, F-38042 Grenoble, France}
\author{S. Raymond}
\affiliation{Universit\'e Grenoble Alpes, CEA, IRIG, MEM, MDN, 38000 Grenoble, France}
\author{E. Ressouche}
\affiliation{Universit\'e Grenoble Alpes, CEA, IRIG, MEM, MDN, 38000 Grenoble, France}
\author{G. Balakrishnan}
\affiliation{Department of Physics, University of Warwick, Coventry, CV4 7AL, United Kingdom}
\author{S. Petit}
\email[]{sylvain.petit@cea.fr}
\affiliation{Laboratoire L\'eon Brillouin, Universit\'e Paris-Saclay, CNRS, CEA, CE-Saclay, F-91191 Gif-sur-Yvette, France}

%
%
\title{Spin dynamics and unconventional Coulomb phase in \ndzr}
%
\begin{abstract}
We investigate the temperature dependence of the spin dynamics in the pyrochlore magnet \ndzr\ by neutron scattering experiments. At low temperature, this material undergoes a transition towards an ``all in - all out'' antiferromagnetic phase and the spin dynamics encompass a dispersion-less mode, characterized by a dynamical spin ice structure factor. Unexpectedly, this mode is found to survive above $T_{\rm N} \approx 300$ mK. Concomitantly, elastic correlations of the spin ice type develop. These are the signatures of a peculiar correlated paramagnetic phase which can be considered as a new example of Coulomb phase. Our observations near \TN\ do not reproduce the signatures expected for a Higgs transition, but show reminiscent features of the ``all in - all out'' order superimposed on a Coulomb phase.
\end{abstract}
%
\maketitle

Geometrical frustration is well known to be one of the key ingredients leading to unconventional states of matter, especially in magnetism \cite{Lacroix11,Gardner10}. Among them, spin ice and more generally Coulomb phases \cite{Henley10} have attracted significant interest. These can be considered as an original state of matter formed by disordered degenerate configurations where local degrees of freedom remain strongly constrained at the local scale by an organizing principle. In the case of spin ice, these degrees of freedom are Ising spins, sitting on the sites of a pyrochlore lattice formed of corner sharing tetrahedra and aligned along the axes which connect the corners of the tetrahedra to their center. The organizing principle, the ``ice rule", states that each tetrahedron should have two spins pointing in and two out, in close analogy with the rule which controls the hydrogen position in water ice \cite{Harris97}. Importantly, the idea that this local constraint can be considered as the conservation law of an ``emergent" magnetic flux (${\bf \nabla} \cdot {\bf B} = 0$) was quickly imposed \cite{Isakov04, Henley05, Castelnovo08}.
Quantum fluctuations can cause this flux to change with time, giving rise to an emergent electric field, and eventually to an emergent quantum electromagnetism \cite{Hermele04, Shannon12, Benton12}. This quantum spin ice state hosts spinon (monopole in the spin ice language \cite{Gingras14}) and photon like excitations. Despite much work, however, experimental evidence for this enigmatic physics remains elusive, with the possible exception of \prhf\ \cite{Sibille18}. Indeed, the conditions for the realisation of this so-called quantum spin ice state are drastic: transverse terms have to be sizable in the Hamiltonian to enable fluctuations out of the local Ising axes, but should remain small enough to prevent the stabilization of classical phases, called Higgs phases, characterized by ordered components perpendicular to these axes \cite{Savary12, Gingras14, Hao14}. 

The pyrochlore material \ndzr\ offers the opportunity to approach this issue. Recent studies suggest that below 1 K this compound hosts a correlated state, which could be a remarkable novel example of Coulomb phase \cite{Petit16, Xu20}. This phase would be described by a ``two in -- two out'' rule as in spin ice,
but built on a pseudospin component different from the conventional $\langle 111 \rangle$ Ising one. The ``all in -- all out'' (AIAO) ordering previously observed  below $T_{\rm N} \approx 300$ mK \cite{Lhotel15, Xu15} would then correspond to the pseudospin ordering in directions perpendicular to the components responsible for the ``high temperature" Coulomb phase. It was proposed that a Higgs mechanism may account for this transition \cite{Xu20}. Such a process is invoked in $U(1)$ quantum spin liquids when the deconfined spinon excitations undergo a Bose-Einstein condensation, resulting in a Higgs phase along with a gapped photon excitation \cite{Pekker15, Savary12, Chang12}.

In this letter, we show that the paramagnetic phase of \ndzr\ does carry elastic spin ice-like correlations, and thus confirm the proposed Coulomb phase picture above \TN. We present a detailed study of the spin dynamics as a function of temperature and explore the nature of this Coulomb phase above and close to the transition. The spin excitations of \ndzr\, deep in the AIAO phase include a peculiar spectrum with a flat band at the energy $E_0\approx$ 70 $\mu$eV characterized by a spin ice-like ${\bf Q}$-dependence \cite{Petit16,Lhotel18,Xu19}. Using neutron scattering experiments, we report the temperature dependence of the gap $E_0$, and reveal that this gap persists above \TN. This result is robust, and withstands a small substitution at the Zr site. The spectra recorded above \TN\ do not show the spinon continuum expected in the Higgs scenario. Instead, we observe dispersive features reminiscent of the AIAO ordered phase superimposed on the Coulomb phase signal. This coexistence suggests that a strong exchange competition is at work in this temperature range, emphasizing the originality of the Coulomb phase above the transition.


The single crystal samples used in this work are the same as in our previous studies (labeled \#1 \cite{Lhotel15, Petit16, Lhotel18} and \#2 \cite{Lhotel18}). 
In addition, results on a single crystal of \ndzrx, with $x=2.4$ \% (Sample \#3) (See supplementary material \cite{supmat}) are presented, not in order to analyze the role of disorder but to illustrate the robustness of the results. 
\nocite{Ciomaga15} \nocite{Ciomaga16} \nocite{Rodriguez-Carvajal93} \nocite{Arnold14} \nocite{Ewings16}
Magnetic properties were measured in very low temperature SQUID magnetometers developed at the Institut N\'eel \cite{Paulsen01}.
The composition and magnetic structure at low temperature were determined using the D23 (CEA-CRG@ILL) neutron diffractometer \cite{supmat}. Polarized neutron scattering experiments were carried out at D7 (ILL) on Sample~\#1. Inelastic neutron scattering (INS) experiments were carried out on the IN5 (ILL) time of flight spectrometer on all samples and on the triple axis spectrometer IN12 (CEA-CRG@ILL) for Sample \#1. The INS data have been analyzed using the {\sc cefwave} software developed at LLB. 

The XYZ Hamiltonian proposed to describe the properties of Nd based pyrochlores due to the peculiar dipolar-octupolar character of the \nd ion \cite{Huang14}, writes:
\begin{equation}
{\cal H} = \sum_{\langle i,j \rangle}\left[{\sf J}_{x} \tau^x_i \tau^x_j + {\sf J}_{y} \tau^y_i \tau^y_j + {\sf J}_{z} \tau^z_i \tau^z_j + {\sf J}_{xz} (\tau^x_i \tau^z_j+\tau^z_i \tau^x_j) \right]
 \label{hxyz}
\end{equation}
In this Hamiltonian, $\tau_i$ is not the actual spin, but a pseudospin which resides on the rare-earth sites of the pyrochlore lattice. Its $z$ component relates to the usual magnetic moment and is directed along the local $\langle 111 \rangle$ directions of the tetrahedra of the pyrochlore lattice. This Hamiltonian can be rewritten by rotating the ${\bf x}$ and ${\bf z}$ axes in the $({\bf x},{\bf z})$ plane by an angle $\theta$. In this $({\bf \tilde{x}},{\bf \tilde{z}})$ rotated frame, the relevant parameters of the Hamiltonian ${\cal H}$ are labeled $\tilde{{\sf J}}_{x,y,z}$, leading to \cite{Huang14, Benton16b}: 
\begin{equation}
\begin{aligned}
{\cal H}_{\rm XYZ} = & \sum_{\langle i,j \rangle}\left[{\tilde {\sf J}_{x}} \tilde \tau^{\tilde x}_i \tilde \tau^{\tilde x}_j +{\tilde {\sf J}_{y}} \tilde \tau^{\tilde y}_i \tilde \tau^{\tilde y}_j + {\tilde {\sf J}_{z}} \tilde \tau^{\tilde z}_i \tilde \tau^{\tilde z}_j \right] \\
& {\rm with} \quad \tan (2 \theta)=\frac{2 {\sf J}_{xz}}{{\sf J}_x - {\sf J}_z}
 \label{hxyz_tilde}
 \end{aligned}
\end{equation} 
With time and maturation of the subject, the estimated parameters for \ndzr\ have evolved. Determinations of the $\tilde{\sf J}_i$ parameters are based on the spin wave spectra measured at very low temperature in zero field \cite{Petit16, Benton16b, Xu19} or applied field \cite{Lhotel18}, while the angle $\theta$ is deduced from the Curie-Weiss temperature \cite{Benton16b} and/or the ordered AIAO magnetic moment \cite{Lhotel18, Xu19}. The sets of reported parameters are summarized in Table \ref{param}, where we have added the parameters refined here for the \ndzrx\ sample (Sample \#3) \cite{supmat} and have revisited the ones of Samples \#1 and \#2. From these values, two interesting features stand out, which remain unexplained to date and should be further explored to ascertain their relevance: (i) the larger the N\'eel temperature, the larger the ordered moment along {\bf z} is. (ii) very similar $\tilde{\sf J}_i$ parameters are obtained for the various samples, despite differences with regard to the amount of impurities or to the ordering parameters.

The ${\sf J}$ parameters lead to an ordered AIAO ground state, where the pseudospins point along the (local) direction ${\bf \tilde{z}}$, turned around the ${\bf z}$-axis towards the ${\bf x}$-axis by the angle $\theta$ \cite{Benton16b}. As shown by INS experiments, peculiar excitations are associated with this ground state. They manifest as an inelastic spin ice like flat mode at an energy $E_0 \approx$ 70 $\mu$eV, above which spin wave branches disperse (See Figure \ref{Figure4}a for Sample \#1) \cite{Petit16}. This excitation spectrum is understood in the framework of the {\it dynamic fragmentation} \cite{Benton16b, Brooks14} as the sum of a dynamic divergence-free contribution, giving rise to the flat mode at $E_0$ and of a dynamic curl-free contribution, which takes the form of the dispersing branches. 
These spin waves correspond to the propagation of magnetically charged excitations and have a spectral weight made of half-moons in reciprocal space \cite{Petit16,Yan18}.

\begin{table*}[!]
\begin{tabularx}{\textwidth}{*{3}{Y}*{4}{P{1.2cm}}Y}
\hline \hline
\multirow{2}{*}{Sample / Ref.} & \multirow{2}{*}{$m_{\rm ord}~(\mu_{\rm B})$} & \multirow{2}{*}{$T_{\rm N}$~(mK)} & \multicolumn{4}{c}{Hamiltonian parameters (K)} & \multirow{2}{*}{$\theta$ (rad)} \\
	 				   &						&		&$\tilde{{\sf J}}_{x}$  & $\tilde{{\sf J}}_{y}$ & $\tilde{{\sf J}}_{z}$ & $\tilde{{\sf J}}_{x}/|\tilde{{\sf J}}_{z}|$ &\\
\hline
\#1	              & $0.8 \pm 0.05$ 	& 285  & 1.18	  &-0.03 	& -0.53 	& 2.20 	&1.23 \\
\#2		& $1.1 \pm 0.1$		& 340	  & 1.0	 	& 0.066 	& -0.5  	&2.0  	&1.09  \\
\#3		            & $1.19 \pm 0.03 $ & 375	  & 0.97	   & 0.21  	& -0.53 	&1.83 	&1.08 \\        
\cite{Xu19} 			     & 1.26			   & 400	  & 1.05	   & 0.16	& -0.53 	&1.98	  &0.98 \\
\cite{Benton16b} 		  & 1.4				   &- 		& 1.2		& 0.0		& -0.55	&2.18	  &0.83 \\
\hline \hline
\end{tabularx}
\caption{Ordered moment $m_{\rm ord}$ along ${\bf z}$, transition temperature $T_{\rm N}$ and Hamiltonian parametrization reported in different studies. $\tilde{{\sf J}}_i$ parameters for Sample \#1 and from Ref. \onlinecite{Benton16b} were obtained from fits of the INS data reported in Ref. \onlinecite{Petit16} and, for Sample \#2 in Ref. \onlinecite{Lhotel18}. $m_{\rm ord}$ from Ref. \onlinecite{Benton16b} is a calculated value. The total Nd$^{3+}$ magnetic moment is estimated to $\approx 2.4~\mu_{\rm B}$ \cite{Lhotel15, Xu15}.}
\label{param}
\end{table*}

\begin{figure}[t!]
\includegraphics[width=8.5cm]{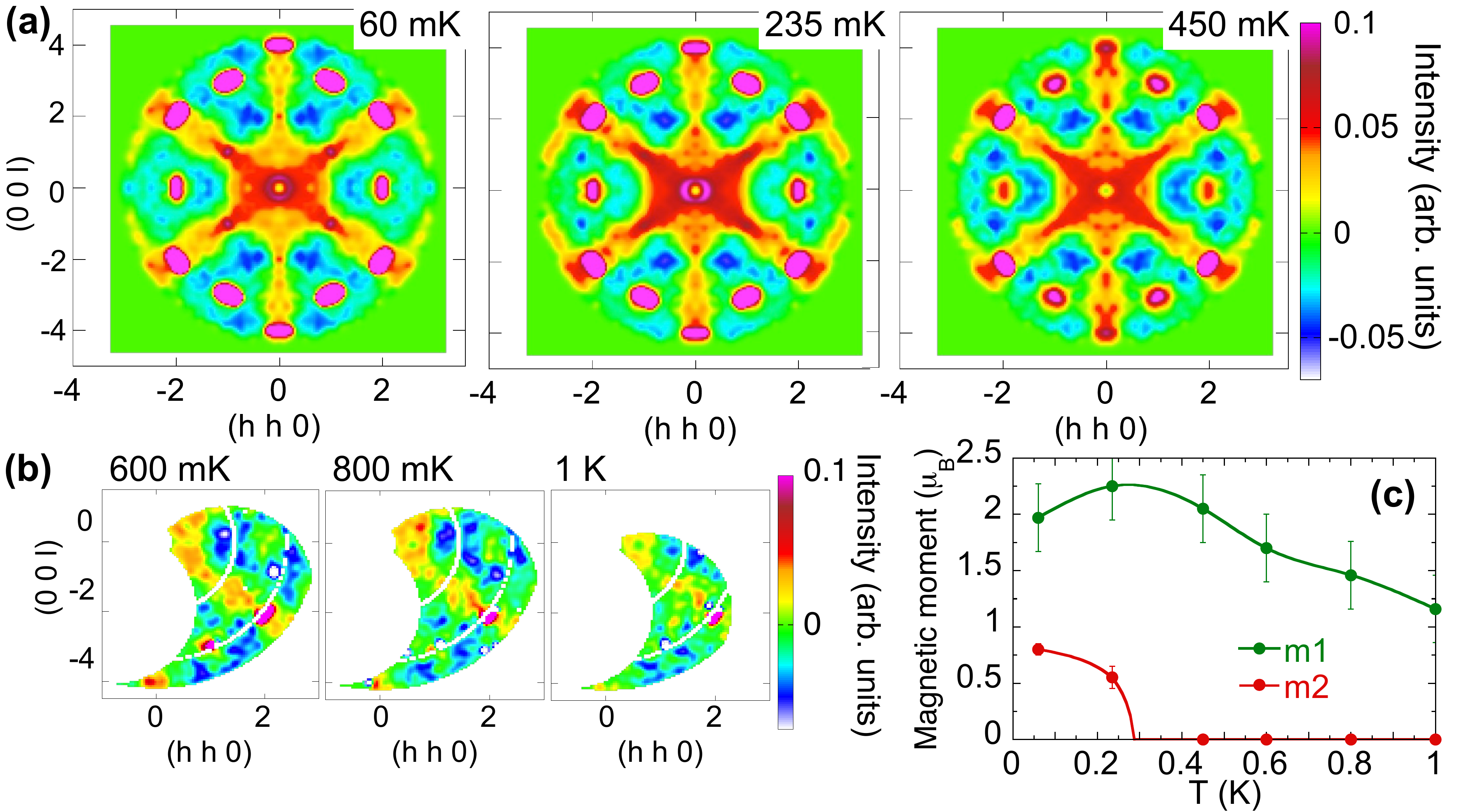}
\caption{\label{Figure1} (a-b) Magnetic instantaneous correlations in Sample~\#1 as a function of temperature. The 10~K dataset has been subtracted as a background reference. Measurements in (a) were symmetrized.  (c) ``Spin ice'' moment $m_1$ and AIAO ordered moment $m_2$ along ${\bf z}$ as a function of temperature  \cite{supmat}. Lines are guides to the eye.} 
\end{figure}

Instantaneous spin-spin correlations $S({\bf Q})$ were measured in Sample \#1 as a function of temperature between 60 mK and 1 K through polarized neutron scattering experiments and are displayed in Figure \ref{Figure1} \cite{supmat}. These measurements integrate over the neutron energy loss up to 3.5~meV, and thus contain both elastic and inelastic signals. At 1~K, a spin ice pattern can barely be observed, revealing the onset of a Coulomb phase. Upon cooling, the spin ice pattern becomes clearly visible below 600~mK. At 450 mK, the magnetic moment $m_1$ responsible for the spin ice-like diffuse scattering is estimated to $2.05 \pm 0.3~\mu_{\rm B}$ \cite{supmat}, to be compared to the 2.4 $\mu_{\rm B}$ full Nd moment \cite{Lhotel15,Xu15}. In addition to this signal, below 800~mK, magnetic diffuse scattering spots appear around $(220)$, $(113)$ and symmetry related positions. Intensity on these positions increases with cooling until they transform into Bragg peaks below \TN\ (285~mK in this sample) characteristic of the AIAO phase. At low temperature, the corresponding ordered magnetic moment is $m_2=m_{\rm ord}= 0.8 \pm 0.05~\mu_{\rm B}$ (from diffraction measurements) and the magnetic contribution to the spin ice like diffuse scattering amounts to $m_1 = 2 \pm 0.3~\mu_{\rm B}$ \cite{supmat} (see Figure \ref{Figure1}c). The moment embedded in the spin ice correlations is thus at maximum around \TN\ and slightly decreases at lower temperature. The diffuse scattering observed in the vicinity of the Bragg peak positions above \TN\ might arise from AIAO diffuse scattering just above the ordering transition, but could also be a signature of deconfined excitations, as proposed in Ref. \onlinecite{Xu20}.

\begin{figure}[t]
\includegraphics[width=8.5cm]{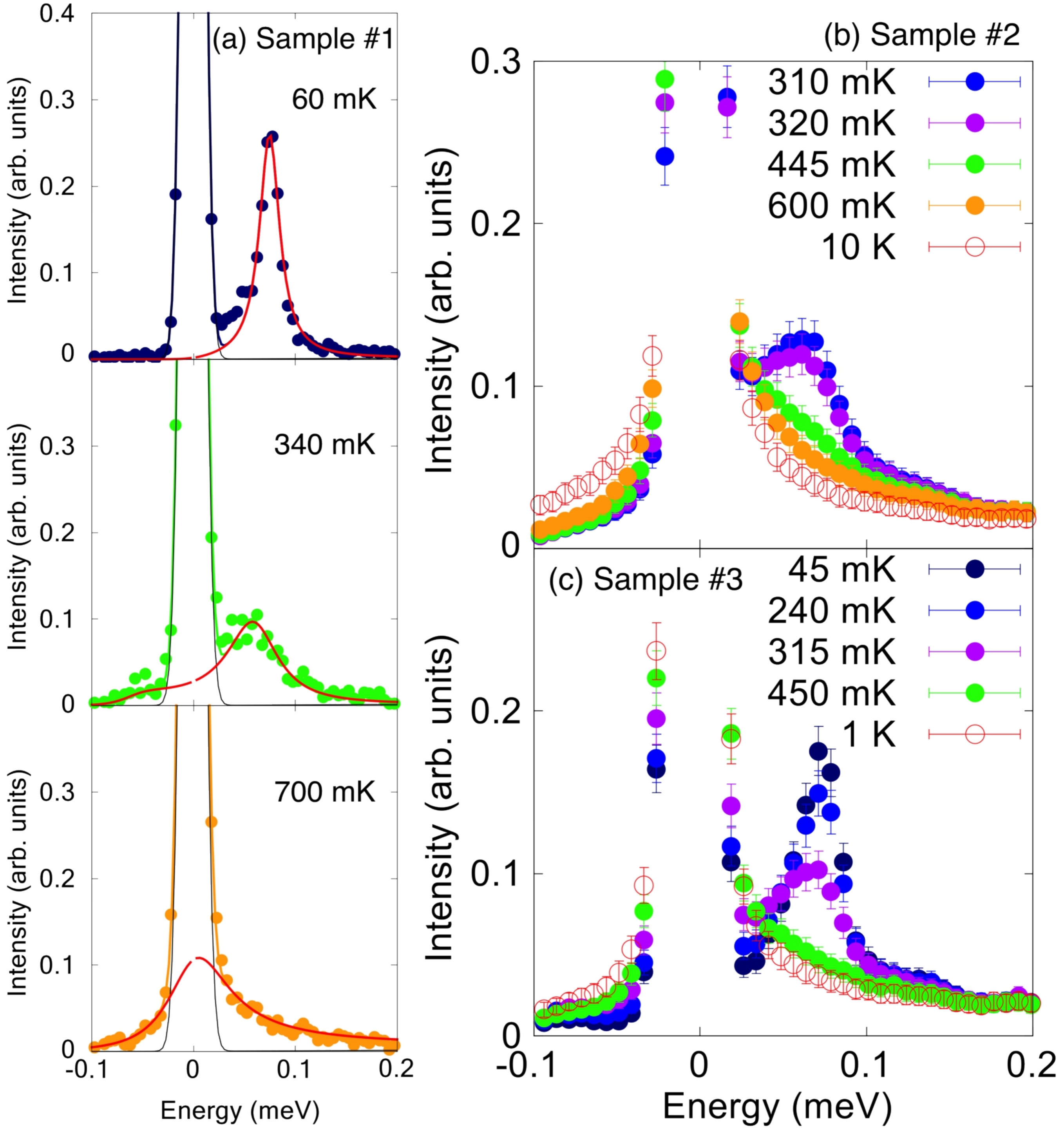}
\caption{\label{Figure2} Spectral function $S(E)$ at different temperatures \cite{supmat} measured at a wavelength $\lambda=$ 8.5 \AA, hence an energy resolution of 20 $\mu$eV: (a) in Sample \#1, integrated around $ {\bf Q}=(0.8~0.8~0.8)$. The grey and red lines correspond to the fitted incoherent elastic $I_c$ and inelastic $S_0$ contributions respectively. 
(b) and (c): integrated over the measured ${\bf Q}$ range in Samples \#2 (b) and \#3 (c). }
\end{figure}

To determine the spectral profile contained in those magnetic correlations, and especially the elastic or inelastic nature of the spin ice correlations associated to $m_1$, INS measurements have been carried out on the three aforementioned samples (see Table \ref{param}) as a function of temperature. To highlight the possible presence of an inelastic flat mode, the ${\bf Q}$-integrated spectral function $S(E)=\int d{\bf Q} S({\bf Q},E)$ was computed.
As this quantity is akin to a density of states, it enhances the contribution of the flat modes contained in the spectrum. Figure \ref{Figure2} displays $S(E)$ at different temperatures. As previously shown \cite{Petit16}, the inelastic flat band is clearly seen at low temperature. It is still visible at finite energy close to \TN\ (320 mK for Sample \#2 and 315 mK for Sample \#3) and above \TN\ (340 mK for Sample \#1), yet broadens significantly upon warming. At the highest temperatures, the signal looks almost quasielastic. To obtain a quantitative insight into the temperature evolution of the mode, data were fitted for the three samples (as shown in Figure \ref{Figure2}a for Sample \#1) to the following model \cite{supmat}: 
\begin{equation}
S(E)= b+I_c(E)+F(E,T) \times \left[ S_0(E)+S_1(E) \right]
\label{eq_S}
\end{equation}
$b$ is a flat background, $I_c(E)$ is a Gaussian function centered at zero energy to account for the elastic incoherent scattering. $F(E,T)=1+n(E, T)$ is the detailed balance factor ($n$ is the Bose-Einstein distribution). $S_0(E)$ and $S_1(E)$ are two Lorentzian profiles, centered on the energy $E_{0,1}$ and of intensity $I_{0,1}$, which represent respectively the flat band and the dispersive mode typical of the spin wave spectrum in \ndzr. 

\begin{figure}[t]
\includegraphics[width=8cm]{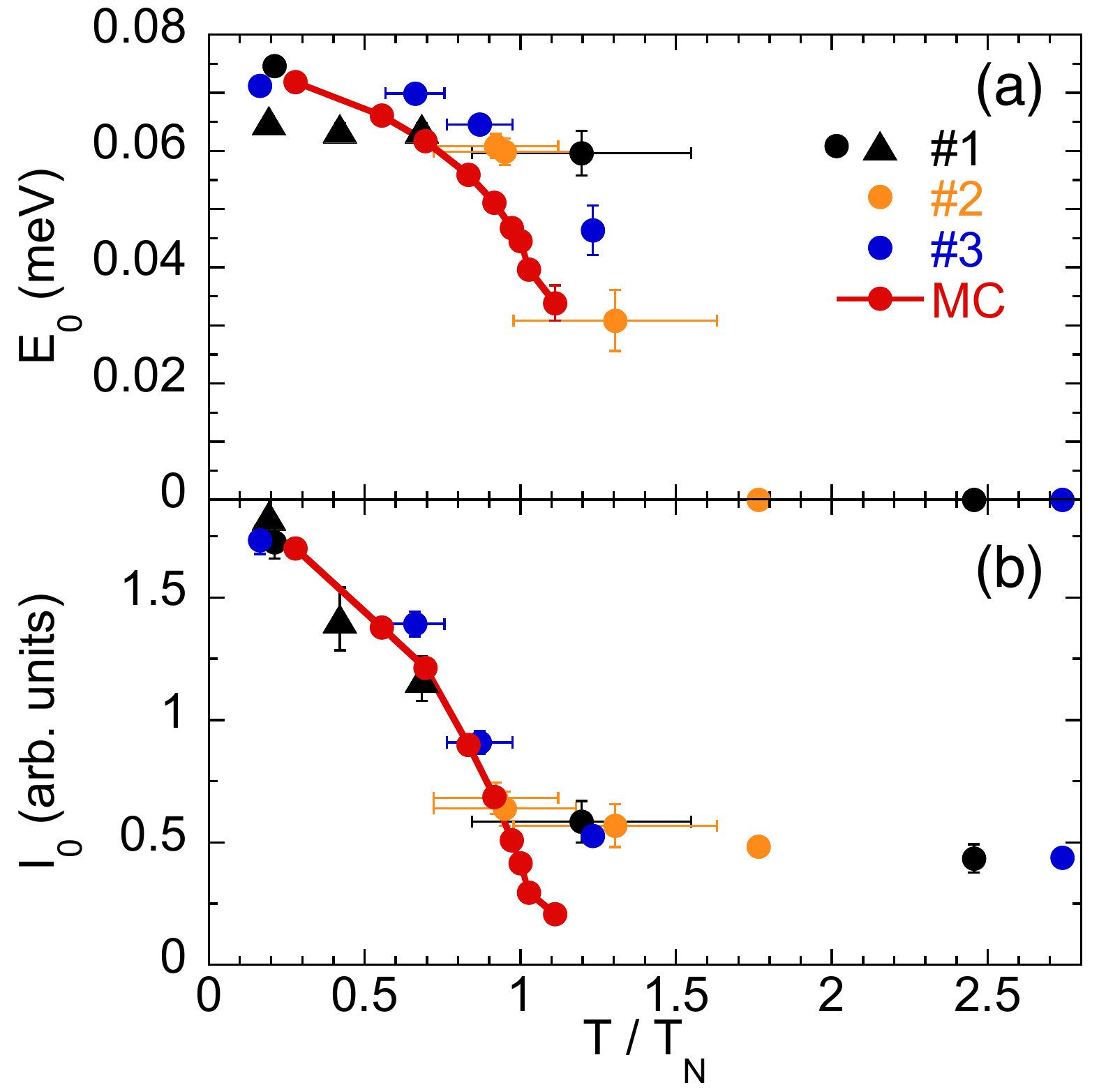}
\caption{\label{Figure3} (a) $E_0$ and (b) $I_0$ as a function of reduced temperature $T/T_{\rm N}$ obtained from measurements on IN5 (dots - see Figure \ref{Figure2}) and IN12 (triangles), together with results from Monte-Carlo (MC) calculations from Ref. \onlinecite{Xu20} (red dots) \cite{supmat}. Lines are guides to the eye. The large $I_0$ experimental value when $E_{0}=0$ is the signature of the persistent quasielastic contribution above $T_{\rm N}$. }
\end{figure}

\begin{figure*}[t!]
\includegraphics[width=18cm]{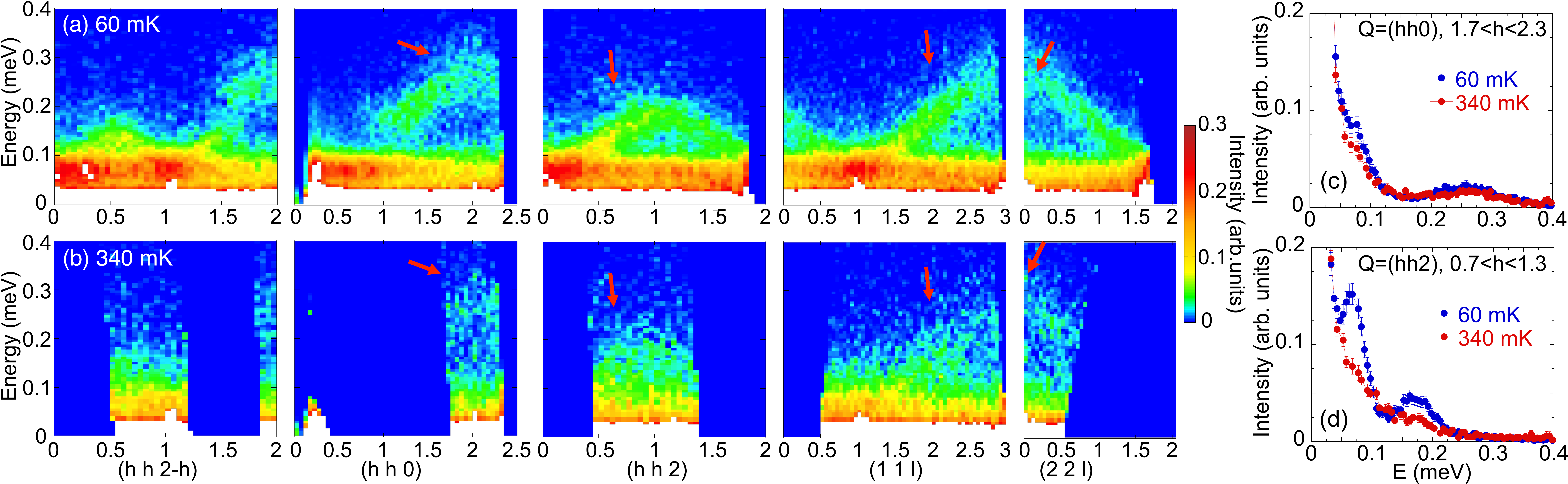}
\caption{\label{Figure4} INS spectra of Sample \#1 along several high symmetry directions at 60 mK (a) and 340 mK (b), measured on IN5 with $\lambda$=6 \AA. Red arrows highlight the dispersive modes and their fingerprints above \TN. (c-d) Constant ${\bf Q}$-cuts at these two temperatures, integrated (c) along $(h h 0)$ and (d) along $(h h 2)$.}
\end{figure*}

The determined positions $E_0$ and intensities $I_0$ are shown in Figure~\ref{Figure3} as a function of the temperature normalized to \TN\ for the three samples. As anticipated from Figure \ref{Figure2}, with increasing temperature, the band at $E_0$ softens and broadens while its intensity decreases. Nevertheless, $E_0$ is non-zero at \TN\ and a persistent dynamical behaviour is observed in all samples at and above \TN, up to about $2 T_{\rm N}$. Finally, the width of the features above the flat mode makes it hard to extract quantitative information from $S_1$. However, close examination of $S({\bf Q},E)$ measured for Sample \#1 above \TN\ at 340 mK (see Figure~\ref{Figure4}) shows that, in all investigated directions, besides a strong quasielastic contribution (the inelastic mode being hardly discernible due to the energy resolution and the color scale), weak features are present close to the position of the low temperature dispersions. These spin wave fingerprints, highlighted by arrows on Figure \ref{Figure4}(b) and which manifest as a broad signal in ${\bf Q}$-cuts (Figure \ref{Figure4}(c-d)), are not compatible with the excitation spectrum expected in the presence of monopole creation and hopping \cite{Xu20}.

Several striking features emerge from these measurements. INS experiments reveal that the intensity $I_0$ of the inelastic spin ice mode decreases when increasing temperature. Since D7 polarized experiments show that the full spin ice correlations, elastic and inelastic, are strongest around \TN, the spin ice pattern observed above \TN\ must contain a new spin ice contribution, likely elastic, and different from the inelastic mode at $E_0$. This is confirmed by magnetization measurements,
which point to ferromagnetic-like correlations, as expected for spin ice \cite{supmat}. This elastic signal could not be directly identified in the elastic line of the IN5 data \cite{supmat} certainly due to background issues, but we should stress that the D7 polarization analysis is definitely the most appropriate way to remove properly nuclear contributions and visualize small magnetic contributions. These results thus point to the coexistence of {\it two} spin ice-like contributions, an elastic and an inelastic one with different origins, and different temperature dependences. 

These two contributions can be understood as the manifestation of the strong competition at play between the pseudospin components of Nd. The negative value of $\tilde{\sf J}_z$ (see Table \ref{param}) promotes an AIAO phase built on ${\tilde \tau}^{\tilde{z}}$ while the positive $\tilde{\sf J}_x$ favors a Coulomb phase, similar to a spin ice phase, but built on ${\tilde\tau}^{\tilde{x}}$. For $\tilde{\sf J}_x/|\tilde{\sf J}_z|\approx2$, the value determined for \ndzr, the former is stabilized at low temperature and the latter at finite temperature, due to the large entropy associated to the Coulomb phase. In these two regimes, spin ice contributions are expected, an {\it elastic} one in the Coulomb phase at ``high" temperature, and an {\it inelastic} one in the AIAO ordered phase (accompanied by dispersive excitations). Remarkably, the observable $\tau_z$, which corresponds to the magnetic dipolar moment along the local $\langle 111 \rangle$ axes, is a combination of the ${\tilde \tau}^{\tilde{x}}$ and ${\tilde \tau}^{\tilde{z}}$ components of the pseudospin. It thus holds the two competing contributions (AIAO and Coulomb), which contrasts with the conventional spin ice case where the $z$ component carries elastic spin ice correlations only.

The present results shed light on the manner in which the system evolves from the ``high" temperature Coulomb phase to the low temperature AIAO ordered phase. At high temperature, around 1~K, the elastic spin ice signal characteristic of the ${\tilde\tau}^{\tilde{x}}$ Coulomb phase appears first. Upon cooling, the inelastic spin ice contribution along with dispersive spin wave branches emerge above \TN\ and coexist with the elastic one. They can naturally be considered as excitations stemming from the short-range AIAO correlations of the ${\tilde \tau}^{\tilde{z}}$ component observed below 800 mK (see Figure~\ref{Figure1}). 

The system enters the long-range AIAO ordered state at a temperature $T_{\rm N} \approx 300$ mK. It corresponds to about $|\tilde{\sf J}_x|/4$, thus to a temperature scale far above the one obtained theoretically for the stabilization of the quantum regime of spin ice, which is estimated to a few percents of the characteristic exchange interaction \cite{Savary13, Huang18}. This indicates that the Coulomb phase remains in its thermal regime down to \TN. Surprisingly, the ordering temperature is larger than semi-classical Monte-Carlo calculations predictions \cite{Xu20}. At \TN, the excitation spectrum is gapped, with the coexistence between the elastic spin ice component and the inelastic spectrum typical of AIAO ordering. The lack of a spinon continuum which would condense at \TN\ seems to preclude a transition driven by a Higgs mechanism.

Deeper in the AIAO phase, the inelastic component - together with the Bragg peaks - develops at the expense of the elastic component. The weak maximum of the spin ice $m_1$ moment around \TN\ can thus be interpreted as due to the rise of the inelastic spin ice mode along with the persistence of the elastic contribution of the Coulomb phase. The coexistence of the elastic and inelastic signals is consistent with MC calculations \cite{Xu20}, even if, close to \TN, the two modes are less distinguishable in the experiments than in the calculations due to the strong broadening of the inelastic mode. Although some distribution is observed between the samples, the measured temperature dependence of the inelastic spin ice mode, described by the energy $E_0(T)$ and intensity $I_0(T)$, is also consistent with calculations \cite{supmat}, despite a slightly stronger inelastic component in experiments above \TN\ (see Figure \ref{Figure3}).

In summary, we find that with increasing temperature, the now well-established flat spin ice band characteristic of the AIAO ground state in \ndzr, softens while its intensity decreases. The energy of this mode remains however finite at and above \TN\ and becomes overdamped with increasing the temperature further. At the same time, a new elastic spin ice component appears. The nature of the correlated phase above \TN\ is thus highly unconventional with the coexistence of an (elastic) Coulomb phase and fragmented excitations, resulting from the competition between the different terms of the Hamiltonian. Our observations support a picture where the AIAO ordering arises from a thermal spin ice phase, a scenario which is well accounted for by semi-classical MC calculations from Ref.~\onlinecite{Xu20}, and is different from the proposed Higgs transition. When increasing the ratio $\tilde{\sf J}_x/|\tilde{\sf J}_z|$, reentrant behaviors are predicted \cite{Xu20} while the system approaches a quantum spin liquid ground state \cite{Benton16b}.
Tuning the parameters of the Hamiltonian (\ref{hxyz_tilde}) with novel materials would thus be of high interest to understand the unusual behavior of \ndzr\ and explore the frontiers between thermal and quantum regimes. 
\acknowledgements
The work at the University of Warwick was supported by EPSRC, UK through Grant EP/T005963/1. M. L. and S.P. acknowledge financial support from the French Federation of Neutron Scattering (2FDN). M. L. acknowledges financial support from Universit\'e Grenoble-Alpes (UGA). M.L., E.L. and S.P. acknowledge financial support from ANR, France, Grant No. ANR-19-CE30-0040-02. S.P. and E.L. acknowledge F. Damay for helpful remarks and J. Xu for providing the data of his calculations. E.L. acknowledges C. Paulsen for the use of his magnetometers. 

\bibliography{biblio_NdZrvsT}	

\clearpage 
\renewcommand{\thefigure}{S\arabic{figure}} 
\renewcommand{\theequation}{S\arabic{equation}} 
\renewcommand{\thetable}{S\arabic{table}}

\setcounter{figure}{0} 
\setcounter{equation}{0} 
 \setcounter{table}{0}

\onecolumngrid

\begin{center} {\bf \large Spin dynamics and unconventional Coulomb phase in \ndzr \\ \medskip Supplementary Material} \end{center}
\vspace{.5cm}

\section{Single crystal growth}
Single crystals of Nd$_2$(Zr$_{1-x}$Ti$_x$)$_2$O$_7$ ($x = 0$ and 0.025) were grown by the floating zone method, using a four-mirror xenon arc lamp optical image furnace \cite{Ciomaga15, Ciomaga16}. A summary of the conditions used for each crystal growth is given in Table \ref*{table_growth}.

\begin{table}[h!]
\begin{tabularx}{\textwidth}{YcYcYc}
\hline \hline
\multirow{2}{*}{Crystal} & \multirow{2}{*}{Sample label} & Lattice parameter & Growth rate & Growth atmosphere,  & Feed / seed  \\ 
				    &						    & \AA                       &(mm/h)         & pressure 
                    &rotation rate (rpm) \\ \hline
\ndzr	& Sample \#1 & $10.66 \pm 0.02$ &12.5 & Air, ambient & 15 / 30 \\		
\ndzr & Sample \#2 & $10.66 \pm 0.04$ & 15 & Air, ambient & 20 / 25 \\	
\ndzrx & Sample \#3 & $10.65 \pm 0.02$ &10 & Air, ambient &15 / 5 \\
\hline \hline
\end{tabularx}
\caption{ \label{table_growth}
Summary of the samples with their crystal growth conditions. The lattice parameters were obtained at 6 K on neutron diffractometers (Sample \#1 and \#3) and triple axis spectrometers (all samples). }
\end{table}

Two different pure \ndzr\ samples had to be used in inelastic neutron scattering experiments, because the first one broke when warming up the dilution fridge after an experiment.


\section{Characterization of the Ti substituted sample (Sample \#3)}
\label{ti-sample}
\begin{figure}[h]
\includegraphics[width=13.5cm]{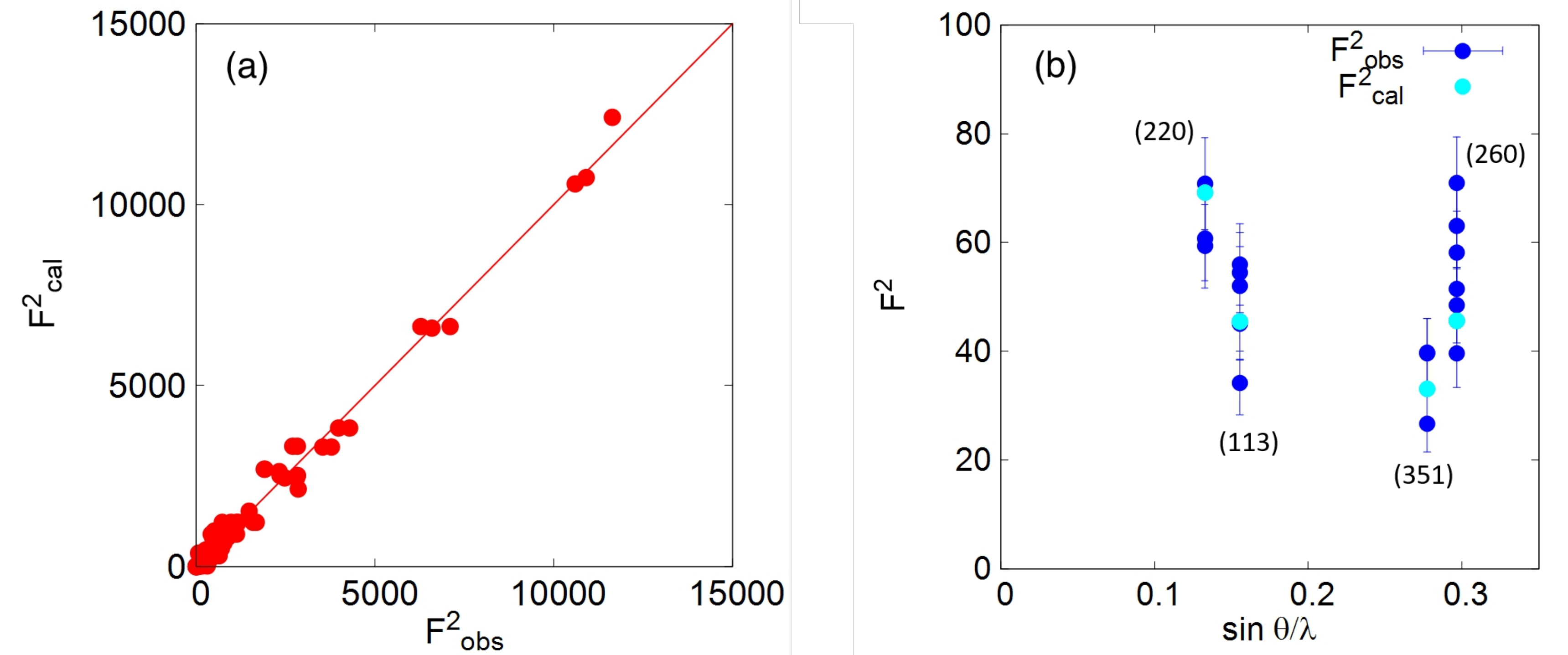}
\caption{\label{ndzrti-D23} (a) Refinement of the crystal neutron structure factors at 6 K, giving a refined Ti content equal to 2.4 \%. (b)~Measured intensity on the magnetic peaks (220), (113), (351) and (260), and symmetry related peaks at 60 mK, obtained from the difference with the 6 K data, and compared to the refined intensity. 
}
\end{figure}

We have studied a substituted sample, in which a small content of Zr is replaced by Ti, slightly shrinking the structure. The nominal composition of the studied sample is 2.5 \% of Ti atoms. As shown below, this substitution only slightly affects the magnetic properties of the magnetic Nd$^{3+}$ sublattice and the low temperature properties are qualitatively the same. 

The value of the Ti content was refined by neutron diffraction, thanks to 
the significant contrast between Zr and Ti. A series of Bragg peak intensities was collected at 6 K on the single crystal neutron diffractometer D23 (CEA CRG-ILL). The data are in agreement with the pyrochlore structure 
($Fd{\bar 3}m$ space group), with a lattice parameter of 10.65 \AA\ and the 48f oxygen atoms at the position $x_{\rm 48f} = 0.336$. The Ti content is found to be 2.4 \%. The {\sc Fullprof} refinement \cite{Rodriguez-Carvajal93} of the structure factor is shown on Fig. \ref*{ndzrti-D23}(a). 

The N\'eel temperature was determined from very low temperature magnetization measurements, and found to be $T_{\rm N}=375$ mK.

The magnetic contribution raises below $T_{\rm N}$ in neutron diffraction 
measurements on top of the crystalline peaks. The Fullprof refinement below $T_{\rm N}$ confirms the same ``all in - all out'' (AIAO) magnetic structure as in the pure sample (Fig. \ref*{ndzrti-D23}(b)). At 60 mK, the refined ordered Nd$^{3+}$ magnetic moment is $1.19 \pm 0.03~\mu_{\rm B}$.


\section{Measurements and analysis of polarized neutron scattering experiments}
\label{D7}
\begin{figure*}[h!]
\includegraphics[width=16cm]{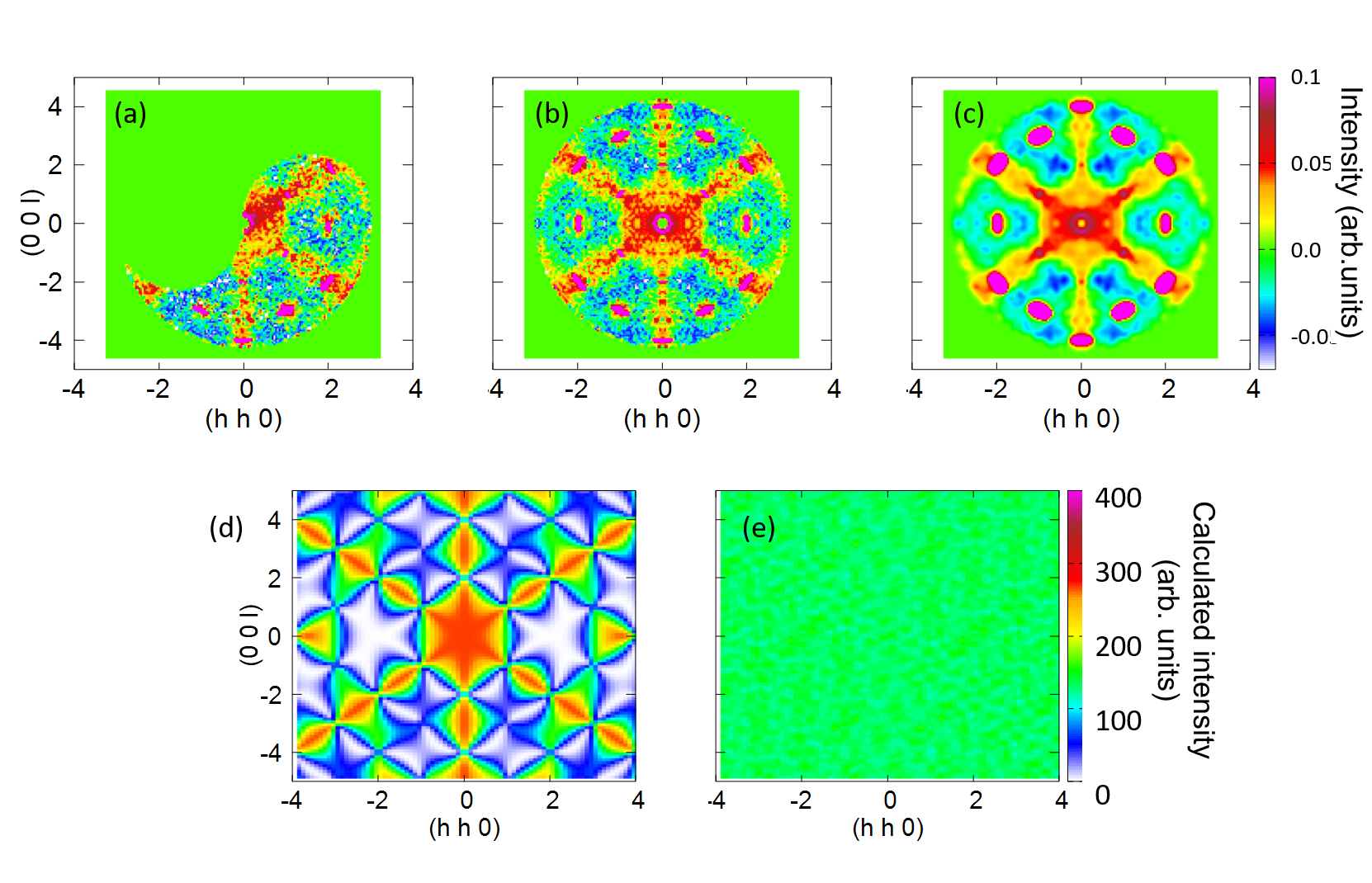}
\caption{\label{App-D7} 
D7 data and analysis. Panels (a,b,c) display respectively the raw data at 
60 mK, the symmetrized and noise filtered data (10 K data have been subtracted). Panel (d) shows the spin ice magnetic scattering function calculated for a lattice containing 5488 spins. Panel (e) shows the same calculation assuming random $\pm 1$ Ising spins. It serves as a background reference, and is subtracted from (d), just as the 10 K data are subtracted from the low temperature data.}
\end{figure*}

In polarized neutron experiments carried out at D7 (ILL, France), we used 
the $P_z$ polarization mode, ${\bf z_Q}$ being the axis normal to the scattering plane and parallel to $[1 \bar{1} 0]$. We measured $N+I^{(z)}$ in the NSF channel, and $I^{(y)}$ in the SF channel. Here we use conventional notations: $N$ is the crystalline structure factor, while $I^{(y)}$ and $I^{(z)}$ denote the spin-spin correlation functions between spin components parallel to ${\bf y_Q}$ and ${\bf z_Q}$ respectively. The ${\bf y_Q}$ axis lies within the scattering plane, perpendicular both to ${\bf Q}$ and ${\bf z_Q}$. We used a wavelength $\lambda = 4.85$ \AA. The sample was rotated by steps of 1 degree, and 2 positions of the detector bank have been combined. Standard corrections (vanadium and quartz) have been processed. Finally, in order to eliminate any background contribution, the data recorded at 10 K have been subtracted from the data taken at lower temperatures.

In Figure 1 of the main article, the data have been symmetrized (for the three lowest temperatures) while the noise was reduced by a mean filtering. This image processing based treatment tends to reduce the variation between one pixel and the next. The idea of mean filtering is to replace each pixel value with the average value of its neighbors, including itself. This has the effect of eliminating pixel values which are unrepresentative of their surroundings. For the sake of illustration, Figure \ref*{App-D7}(a-c) shows the different steps of this processing for the 60 mK data.
\bigskip

Unfortunately, it was not possible to determine intensities in absolute units from the D7 measurements. To determine the magnetic moment responsible for the spin ice-like diffuse scattering, we had to proceed in an alternative manner. To this end, we carried out a series of calculations, assuming a ``theoretical sample crystal'' consisting of Ising spins (of length unity) located at the rare earth sites of a pyrochlore lattice of size 
$L$. We have considered $n$ spin ice configurations generated on this pyrochlore lattice (via a Monte-Carlo algorithm) and computed the average structure factor from the obtained magnetic moment ${\bf m}_{i,a=x,y,z}$ at each site $i$. The total magnetic neutron intensity, proportional to the spin-spin correlation, is calculated as: $$I^{(y)}+I^{(z)} = \sum_{i,j} \sum_{a,b=x,y,z} m_{i,a} \left(\delta_{ab}- \frac{{\bf Q}_a{\bf Q}_b}{{\bf Q}^2}\right) m_{j,b}~e^{i {\bf Q}.({\bf R}_i-{\bf R}_j)}$$ and the intensity in the SF $P_z$ mode, corresponding to the data, is given by $\displaystyle I^{(y)} = \sum_{i,j} {\bf m}_{i}.{\bf y}_{\bf Q}\ {\bf m}_j.{\bf y}_{\bf Q}\ e^{i {\bf Q}.({\bf R}_i-{\bf R}_j)}$.

Noteworthy, the Monte-Carlo sampling was checked using the analytical method proposed by C.L. Henley \cite{Henley05}. The spin-spin correlation function (per unit cell) is written as:
\begin{equation*}
I^{(y)} = 4~t_0^2~\sum_{a=x,y,z} M^T_a \left[ I - E(E^+E)^{-1}E^+ \right] M_a
\end{equation*}
$E$ is a 2-column matrix and $M_{a=x,y,z}$ is a collection of 1-column vectors containing the coordinates $a=x,y,z$ of the four magnetic moments ${\bf m}_i$ belonging to a given tetrahedron:
\begin{align*}
E & = 
\begin{pmatrix} 
e^{-i \pi {\bf Q.u_1}} & e^{i \pi {\bf Q.u_1}} \\
e^{-i \pi {\bf Q.u_2}} & e^{i \pi {\bf Q.u_2}} \\
e^{-i \pi {\bf Q.u_3}} & e^{i \pi {\bf Q.u_3}} \\
e^{-i \pi {\bf Q.u_4}} & e^{i \pi {\bf Q.u_4}} \\
\end{pmatrix} \\
M_a&=
\begin{pmatrix}M_{1,a} \\ M_{2,a} \\M_{3,a} \\ M_{4,a} \end{pmatrix}
\quad {\rm where} \quad M_{i,a}=m_{i,a} - \frac{{\bf m}_{i}.{\bf Q}}{Q^2}Q_a -\left(\sum_b \left(m_{i,b} - \frac{{\bf m}_i.{\bf Q}}{Q^2}Q_b\right).z_{{\bf Q},b}\right)z_{{\bf Q},a}  \quad {\rm for} \quad {i=1,2,3,4}
\end{align*}
The four moments are defined as ${\bf m}_1=(c,c,c)$, ${\bf m}_2=(-c,-c,c)$, ${\bf m}_3=(-c,c,-c)$, ${\bf m}_4=(c,-c,-c)$ with $c=1/\sqrt{3}$ and attached to a tetrahedron. The ${\bf u_i}$ are vectors pointing towards the four corners of a tetrahedron: ${\bf u_1}=(d,d,d)$, ${\bf u_2}=(-d,-d,d)$, ${\bf u_3}=(-d,d,-d)$, ${\bf u_4}=(d,-d,-d)$ and $d=1/4$. Proper normalization condition imposes $t_0^2=2$.

In the same way, we computed the structure factor $I^{(y)}_{\rm rd}$ assuming that the Ising spins have purely random $\pm 1$ values. This quantity was used as a background reference, just as the 10~K measurement was used as described above. We eventually considered the case where the spins are arranged in an ``all in -- all out" (AIAO) ordering, leading to magnetic Bragg peaks with a structure factor denoted hereafter $I^{(y)}_{\rm AIAO}$. 

First we calculated the integrated intensity around $(11\bar{3})$ from the experimental data at different temperatures, yielding $I^{\rm exp}_{\rm AIAO}(T)$. On the other hand, the same quantity was determined from $I^{(y)}_{\rm AIAO}$, yielding  $I^{\rm c}_{\rm AIAO}$. Since the actual value of the ordered AIAO moment at low temperature $m_{\rm AIAO}$ (called $m_2$ in the main text)
is precisely known from diffraction measurements (D23), we introduced the normalization factor $c$:
\begin{equation*}
c = \frac{I^{\rm exp}_{\rm AIAO}}{m_{\rm AIAO}^2~I^{\rm c}_{\rm AIAO}}
\end{equation*}
In a second step, we have computed the experimental integrated intensity within a box delineating the arm along $(hh{\bar h})$, yielding  $\Delta I^{\rm exp}_{\rm arm}$. To obtain an estimate of the moment $m_{\rm SI}$ (called $m_1$ in the main text)
involved in the spin ice component, which reflects the evolution seen on the maps presented in Figure 1 of the main text, we proposed to compare $\Delta I^{\rm exp}_{\rm arm}$ to $\Delta I^{\rm c}_{\rm arm} = m_{\rm SI}^2 (I^{(y)}-I^{(y)}_{\rm rd})$.
The estimation is then made quantitative by looking for $m_{\rm SI}$ such that : 

\begin{equation*}
\Delta I^{\rm exp}_{\rm arm}(T) =  c \times  \Delta I^{\rm c}_{\rm arm}
\end{equation*}
The obtained values are listed in Table \ref*{m1m2}. 

Other calculation methods have been tested and lead to similar results in terms of absolute values and evolution with temperature. Furthermore, using $m_{\rm AIAO} = 0.8 \pm 0.05~\mu_{\rm B}$, the accuracy on $m_{\rm SI}$ is estimated to $\pm 0.3~\mu_{\rm B}$. This analysis confirms that the spin ice pattern has a weak maximum close to \TN\ and persists up to 600 mK, i.e. far above \TN. \\

\begin{table*}[!h]
\begin{tabularx}{7cm}{cYcYcY}
\hline
\hline
$T$ (mK) & $I^{\rm exp}_{\rm arm}(T) $  & $m_{\rm SI}$ ($\mu_{\rm B}$) \\
\hline
60    & 0.027 & 1.97 \\
235  & 0.035 & 2.25 \\
450 & 0.030 & 2.05 \\
600 & 0.020 & 1.70 \\
800 & 0.015 & 1.46 \\                    
1000  & 0.0095 & 1.16 \\
\hline
\hline
\end{tabularx}
\caption{\label{m1m2} Values of the spin moment $m_{\rm SI}$ {\it vs} temperature, determined from the procedure described in the text. Note that the temperature of 235 mK was estimated from the amplitude of the magnetic Bragg peaks (the thermometer indicated 300~mK). For 450 mK, no precise determination of the sample temperature could be done, but, in the absence of magnetic Bragg peaks, the temperature was definitely above \TN. }
\end{table*}


\section{Time of flight inelastic scattering measurements}

Inelastic neutron scattering experiments were carried out on the IN5 disk 
chopper time of flight spectrometer (ILL, France). A good compromise between flux, energy resolution and accessible ${\bf Q}$ space was obtained with a wave length $\lambda=6$ or 6.5 \AA. However, to ensure a better energy resolution $\Delta E= 20~\mu$eV, necessary to fully resolve the dynamic spin ice mode at $E_0$, experiments were also conducted with $\lambda=8.5$ \AA. The data were processed with the {\sc Mantid} \cite{Arnold14} and {\sc horace} \cite{Ewings16} softwares, transforming the recorded time of flight, sample rotation and scattering angle into energy transfer and ${\bf Q}$-wave vectors. The offset of the sample rotation was determined based on the Bragg peak positions. In all the experiments, the sample was rotated in steps of 1 degree and the counting time was about 10 minutes per sample position. 

It should be noticed that a very long thermalization time was systematically necessary to cool down the sample to the lowest temperature. In addition, we realized that when warming up from the lowest temperature, the sample temperature was not necessarily the same as the temperature indicated by the thermometer. For this reason, when possible (depending on the ratio between the resolution and the temperature), we have refined the ``true" temperature by fitting the negative energy part of the spectra. It leads to the temperatures indicated on Figures 2, 3 and 4 of the main text, 
which are quite different from the thermometer temperatures. These temperatures are summarized in Table \ref*{table_T}.
 
 \begin{table}[b]
\begin{tabular}{C*{2}{P{3.5cm}}}
\hline
\hline
\multirow{2}{*}{Sample}  &Thermometer & Estimated \\
	& temperature & temperature \\
\hline
\hline
Sample \#1 &	450 mK	  		&		$341 \pm 100$ mK \\ \hline
\multirow{3}{*}{Sample \#2} & 60 mK			& $323 \pm 78$ mK	\\
					& 300 mK			& $313 \pm 68$ mK	\\
					&450 mK 			& $444 \pm 111$ mK \\ \hline
\multirow{2}{*}{Sample \#3} & 275 mK		& $242 \pm 35$ mK	\\
					&350 mK 			& $317 \pm 38$ mK \\					
\hline
\hline
\end{tabular}
\caption{Estimated effective temperatures in the different experiments performed on IN5.}
\label{table_T}
\end{table}

\medskip

\begin{figure}[h!]
\includegraphics[width=8cm]{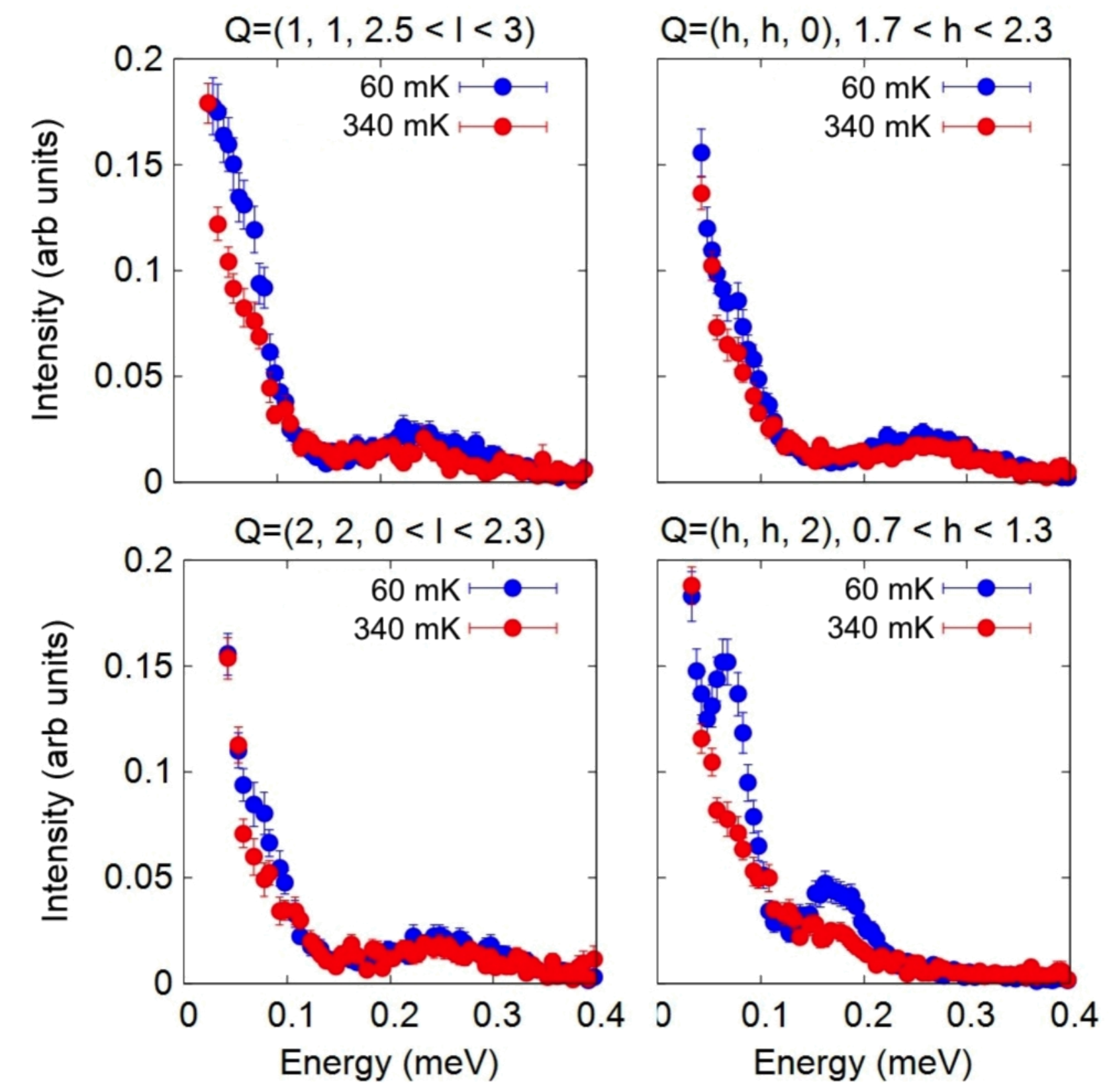}
\caption{\label{coupe} Constant ${\bf Q}$-cuts at two temperatures, the base temperature of 60 mK (blue) and above \TN\ (red) and which clearly show the vestiges of spin waves.}
\end{figure}
Constant ${\bf Q}$-cuts from the data have been performed at the base temperature (typically 60 mK) and above \TN\ (340~mK), to clearly show the persistence of the spin wave signal above \TN. These cuts, displayed in Figure \ref*{coupe} are along $(1,1,\ell)$, $(2,2,\ell)$, $(h,h,0)$ and $(h,h,2)$. 
Two of them are reproduced in the main text. 

This residual spin wave signal above \TN\ is not expected in conventional 
three dimensional paramagnets, in the absence of magnetic frustration. It 
thus would not be observed in a standard ``all in -- all out" antiferromagnet, which is predicted to behave classically close to the antiferromagnetic transition. The persistence of the spin wave signal (and of ``all in 
-- all out" diffuse scattering) quite far above \TN\ in \ndzr\ thus points out the unconventional nature of the magnetism in this compound and is likely related to the strong competition at play with the Coulomb phase observed above \TN.  


\section{Inelastic scattering measurements on a triple axis spectrometer}
\label{tas}

The temperature dependence of the spin dynamics in Sample \#1 was also investigated on the cold TAS spectrometer IN12 (ILL, France). Scans at specific ${\bf Q}$ positions (0.5 0.5 2), (0 0 2.5) and (1.8 1.8 0) have been 
performed at different temperatures ranging from 50 up to 800 mK. Those positions were chosen since they probe different regions with respect to the dispersion. (0.5 0.5 2) essentially probes the flat spin ice band, (1.8 1.8 0) is sensitive to the zone boundary dispersive spin wave mode and (0.5 0.5 2) is somehow intermediate. A final wave vector $k_f=1.05$ \AA$^{-1}$ was used (in combination with nitrogen cooled Be filter) to ensure 
the best energy resolution, $\Delta E=50~\mu$eV. A magnetic field was also applied along $[1\bar{1}0]$. After correction from the detailed balance factor, we computed the difference between data taken a given temperature $T$ and the 800 mK data. Where applicable, we subtracted the data obtained at the same temperature but under a 1 T magnetic field. We could then extract the energy and intensity of the inelastic mode in the same way 
as for TOF measurements. The temperatures below $T_{\rm N}$ were estimated from the intensity of the (220) magnetic Bragg peak.

\section{Spin dynamics in the Ti substituted sample (Sample \#3)}
\label{spindyn-ti}
\subsection{Determination of the parameters}

\begin{figure*}[t!]
\includegraphics[width=16cm]{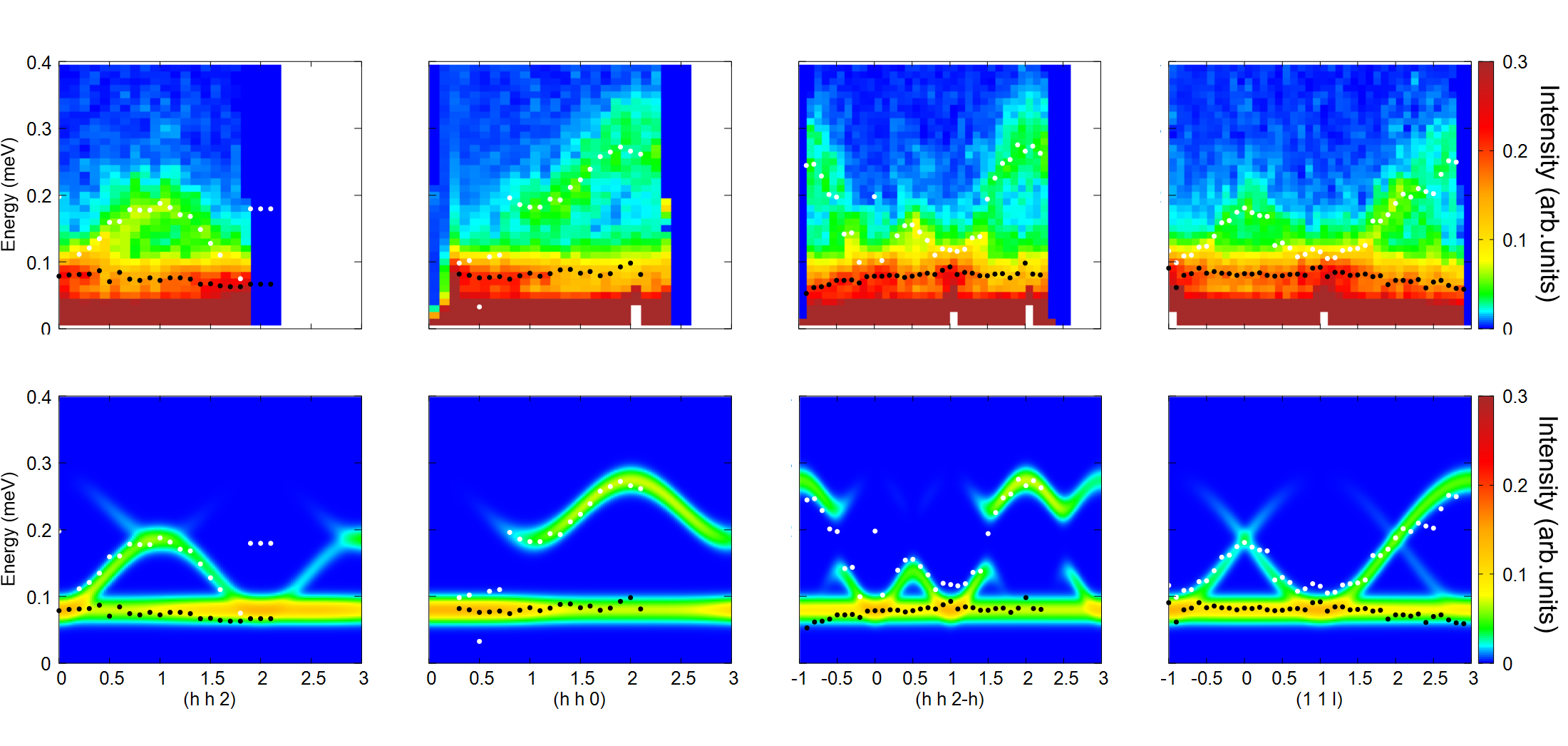}
\caption{\label{App-ins} Top: INS data taken at IN5 at 45 mK on the Sample \#3 along high symmetry directions. Black and white dots are the energies $E_0$ and $E_1$ respectively, fitted according to the procedure described in the main text (see also equation (\ref*{fit_INS}). Bottom: Spin wave calculations performed with the parameters given in Table 1 (main text).}
\end{figure*}

Inelastic neutron scattering data carried out at IN5 (ILL) on a single crystal sample show little evolution compared to the pure sample. The inelastic flat spin ice mode is observed at $E_0 \approx 70~\mu$eV, while the dispersing mode stemming from the pinch point positions unfolds towards the zone centers, for instance $(220)$ or $(113)$. This is illustrated in Figure \ref*{App-ins}, which shows the dispersion along several reciprocal 
directions at 45 mK. 

To determine the parameters of the XYZ Hamiltonian 
\begin{equation*}
{\cal H}_{\rm XYZ} = \sum_{\langle i,j \rangle}\left[ {\tilde {\sf J}_{x}} \tilde \tau^{\tilde x}_i \tilde \tau^{\tilde x}_j +{\tilde {\sf J}_{y}} \tilde \tau^{\tilde y}_i \tilde \tau^{\tilde y}_j + {\tilde {\sf J}_{z}} \tilde \tau^{\tilde z}_i \tilde \tau^{\tilde z}_j \right]
\end{equation*}
(see also equation (2) of the main text), we use analytic calculations giving the energy of the spin ice band \cite{Benton16b}: 
\begin{equation*}
E_0 = \sqrt{(3 |\tilde{{\sf J}}_{z}|-\tilde{{\sf J}}_{x})(3 |\tilde{{\sf J}}_{z}|-\tilde{{\sf J}}_{y})}
\end{equation*}
as well as the energy of the dispersive modes at some high symmetry ${\bf 
Q}$ vectors \cite{Xu19}: 
\begin{eqnarray*}
{\bf Q}=(110), (112) &,~ & \Delta_2 =\sqrt{(3 |\tilde{{\sf J}}_{z}|+\tilde{{\sf J}}_{x})(3 |\tilde{{\sf J}}_{z}|+\tilde{{\sf J}}_{y})} \\
{\bf Q}=(220), (113) &,~ & \Delta_3=3\sqrt{( |\tilde{{\sf J}}_{z}|+\tilde{{\sf J}}_{x})(|\tilde{{\sf J}}_{z}|+\tilde{{\sf J}}_{y})}
\end{eqnarray*}

Simulations have  then been performed to reproduce the data with the {\sc 
cefwave} software developed at LLB using the Hamiltonian (1): 
\begin{equation*}
{\cal H} = \sum_{\langle i,j \rangle}\left[{\sf J}_{x} \tau^x_i \tau^x_j + {\sf J}_{y} \tau^y_i \tau^y_j + {\sf J}_{z} \tau^z_i \tau^z_j + {\sf J}_{xz} (\tau^x_i \tau^z_j+\tau^z_i \tau^x_j) \right]
 \label{hxyz}
\end{equation*}
The ground state configuration is first determined by solving this Hamiltonian at the mean field level, where the expectation values $\langle \tau^{x,y,z}_j \rangle$ are determined in a self-consistent manner. Spin wave calculations are performed using a generalized susceptibility approach out of the obtained configurations. Finally, the neutron cross section is calculated from $\tau^z_i \tau^z_j$ correlations. Notably, the simulations 
performed with the parameters of Table 1 (main text) reproduce quite well 
the data, as shown in Figure \ref*{App-ins}. 

\subsection{Temperature dependence of the spin dynamics}
Inelastic data in Sample \#3 were fitted in the whole measured ${\bf Q}$ space using the model described in the main text: 
 \begin{equation}
S({\bf Q},E)= b+I_c(E)+F(E,T) \times \left[ S_0(E)+S_1(E) \right]
\label{fit_INS}
\end{equation}
$b$ is a flat background (wavelength dependent), $I_c$ is a Gaussian function centered at zero energy to represent the elastic incoherent scattering. $F(E,T)=(1+n(E))$ is the detailed balance factor, and $S_0$ and $S_1$ are two Lorentzian profiles which represent respectively the flat band 
and the dispersive mode typical of the spin wave spectrum in \ndzr. 

\begin{figure*}[h!]
\includegraphics[width=16cm]{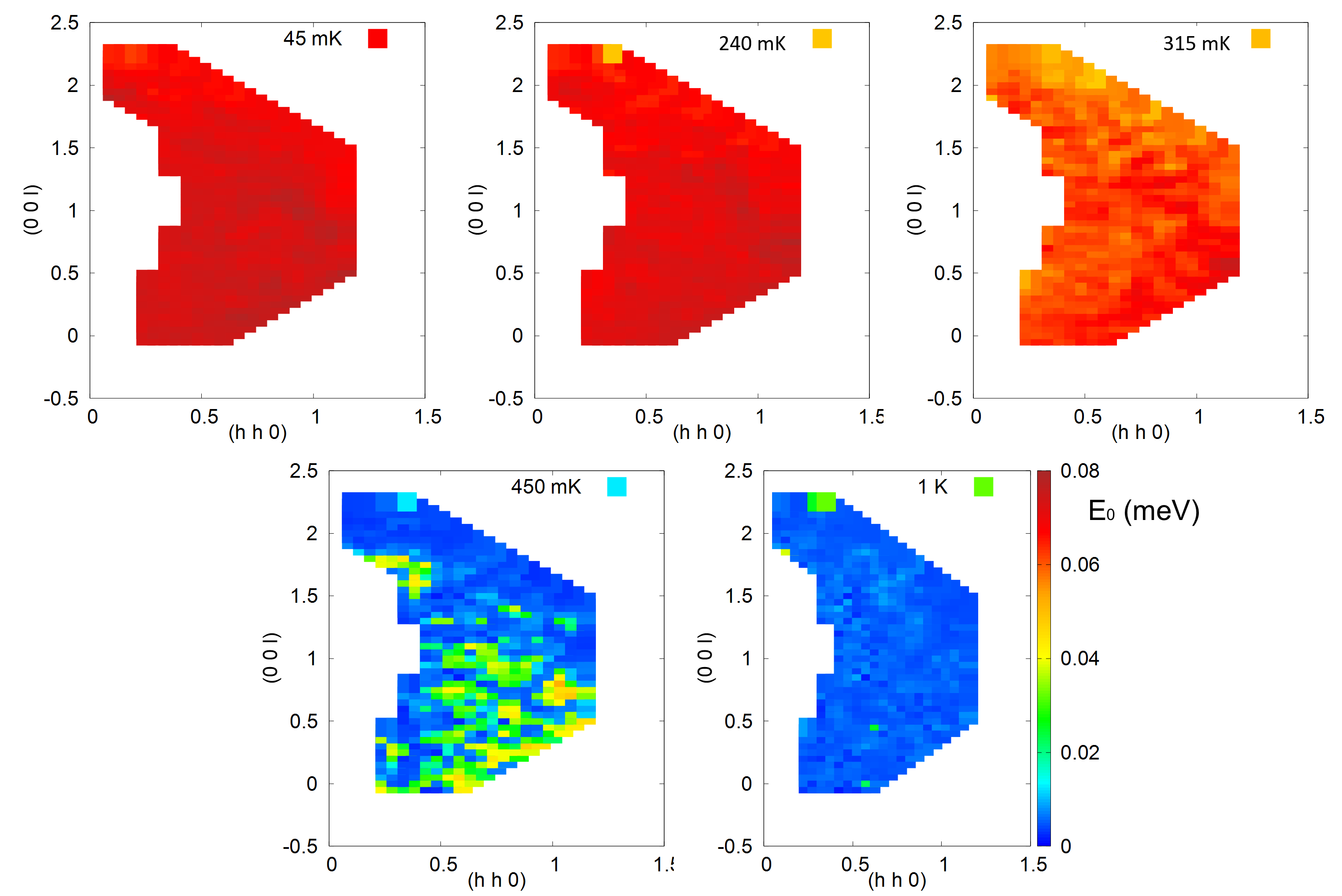}
\caption{\label{App-ins2} Temperature dependence of the flat spin ice band at $E_0$ deduced from the fit in Sample \#3, as described in the main text. The portion of $({\bf Q},E)$ space corresponds to the sector probed by TOF measurements with $\lambda$=8.5 \AA.}
\end{figure*}

\begin{figure*}[h!]
\includegraphics[width=16cm]{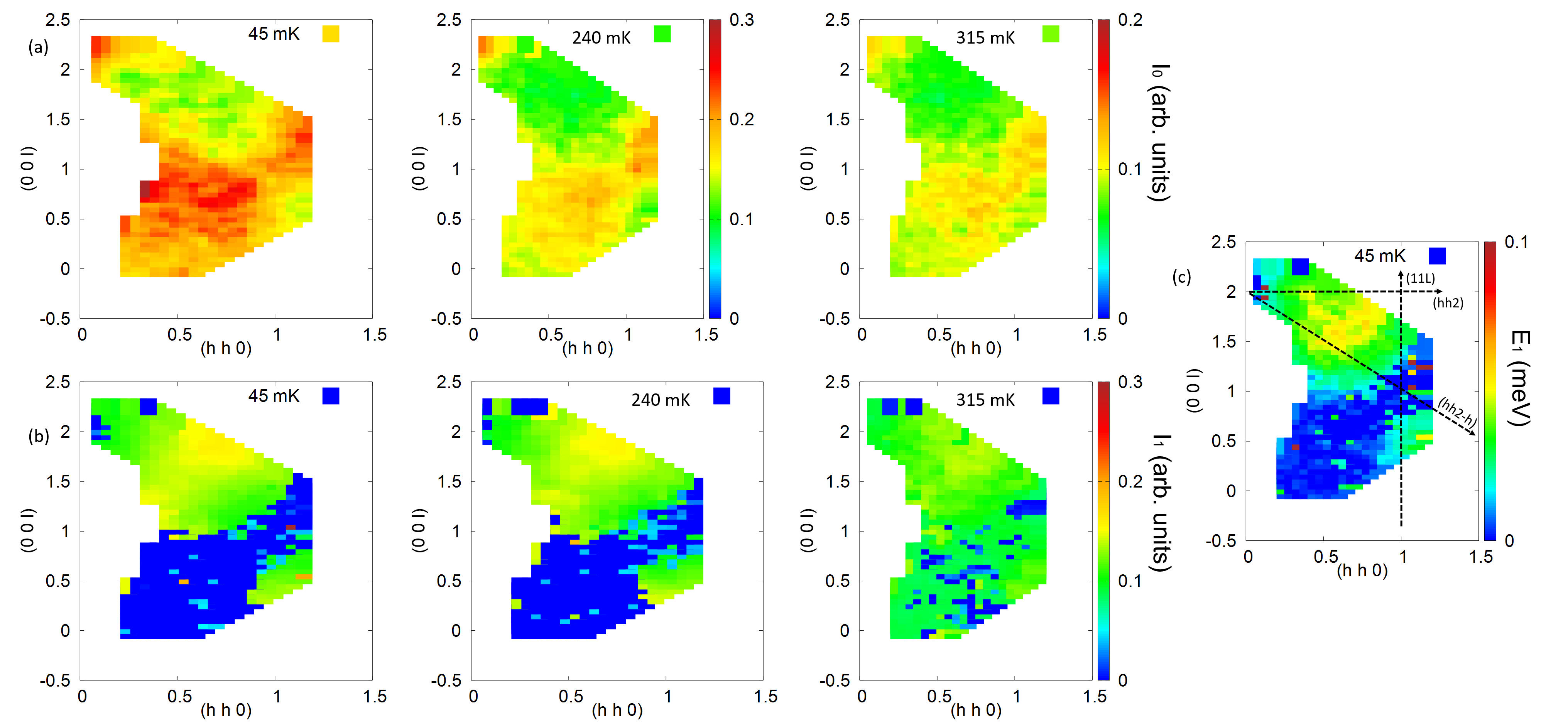}
\caption{\label{App-ins3} (a) Temperature dependence of the intensity $I_0$ of the flat spin ice band from the fit in Sample \#3, as described in the main text. (b) shows the intensity $I_1$ of the dispersive mode, and (c) shows its energy $E_1$. Dashed lines correspond to the directions of the cuts shown in Figure \ref*{App-ins}. The portion of $({\bf Q},E)$ space corresponds to the sector probed by TOF measurements with $\lambda=8.5$~\AA. }
\end{figure*}

Figure \ref*{App-ins2} shows the energy $E_0$ of the flat band at different temperatures in the form of a map over the sector probed by TOF measurements. Figure \ref*{App-ins3} displays the intensities $I_0$ (panel a, upper row) and $I_1$ (panel b, lower row). The map on the right of the same figure shows the energy $E_1$ of the dispersive spin wave mode. To check the overall consistency of the fitting procedure, dashed lines visualize the directions of the scans reported in Figure \ref*{App-ins}. 

Note that for Sample \#1, the fit was carried out at selected ${\bf Q}$ values (${\bf Q} = (0.8\ 0.8\ 0.8)$, $(1.1\ 1.1\ 1.1)$, $(1/2\ 1/2\ 1/2)$, $(1/2\ 1/2\ 3/2)$ and $(3/4\ 3/4\ 3/2)$).

\newpage

\begin{figure}[h!]
\includegraphics[width=8.5cm]{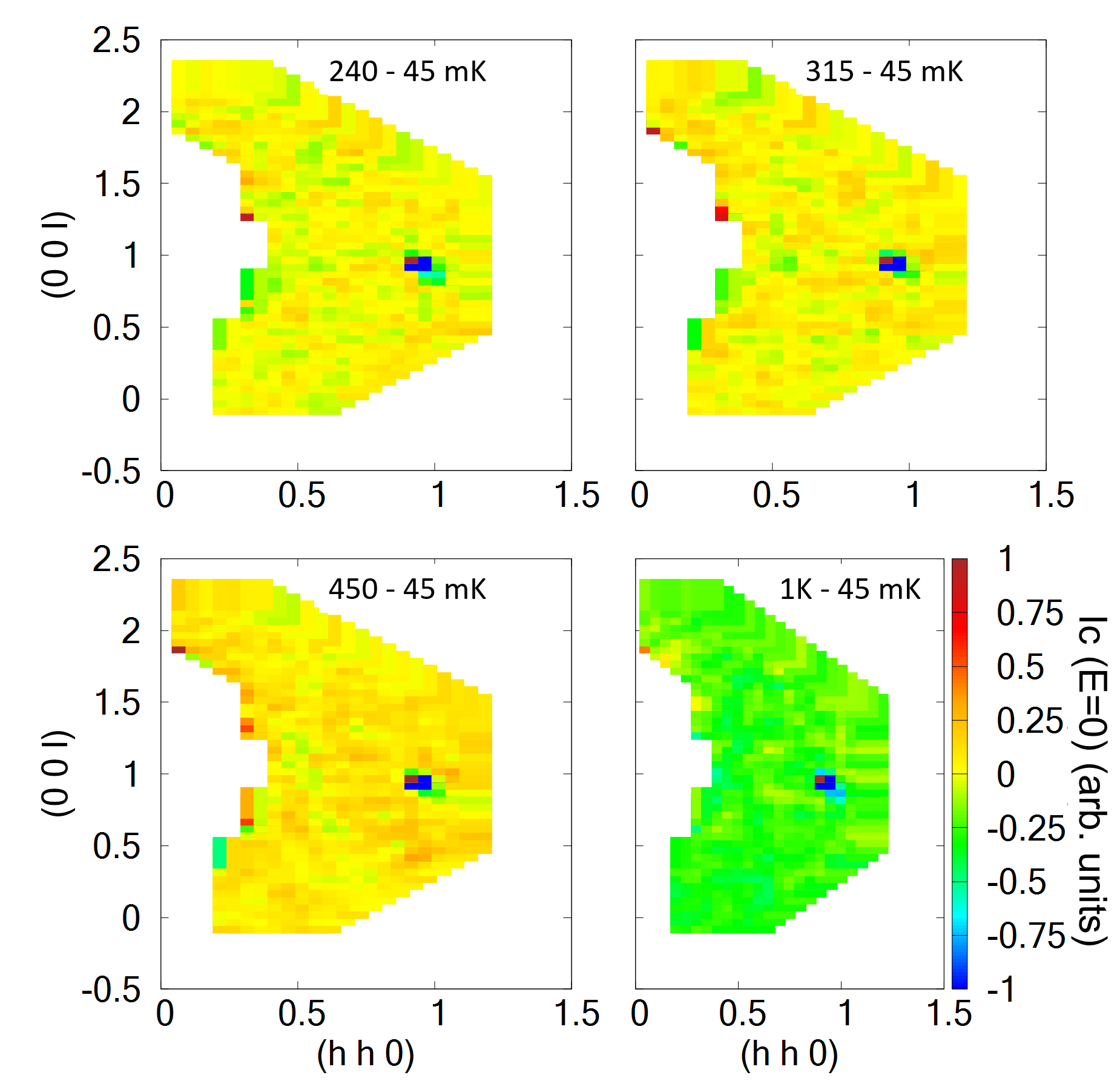}
\caption{\label{App-ins4} Temperature dependence of the intensity of the incoherent scattering $I_c(E=0)$ in Sample \#3. The portion of $({\bf Q},E)$ space corresponds to the sector probed by TOF measurements with $\lambda=8.5$~\AA. }
\end{figure}

 Finally, aiming at identifying a possible elastic contribution with the spin ice structure factor, Figure \ref*{App-ins4} shows the temperature evolution of the intensity of the incoherent elastic contribution, i.e.~the dominant contribution $I_c(E=0)$ in the spectrum, to which the 45 mK map was subtracted. Within experimental uncertainties, these maps are featureless and no spin ice pattern can be clearly distinguished.

\FloatBarrier

\section{Analysis of classical dynamics results}
\label{Benton}

\begin{figure*}[h]
\includegraphics[width=16cm]{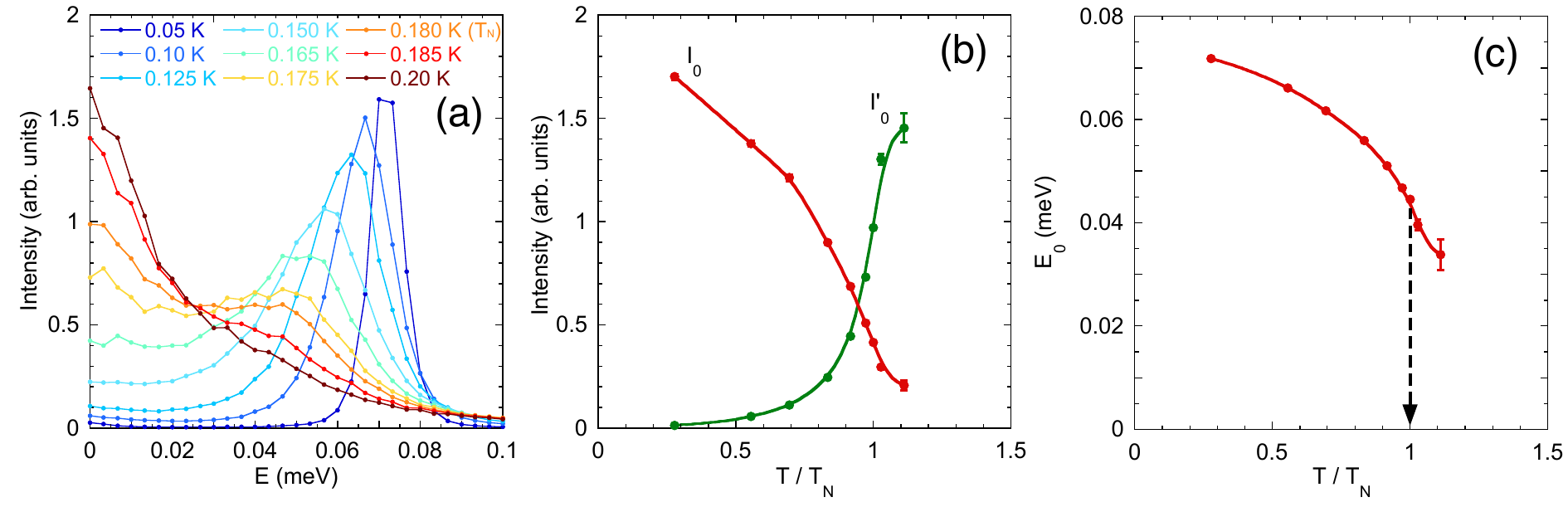}
\caption{\label{App-Benton} (a) From Ref. \onlinecite{Xu20} (courtesy of J. Xu): Evolution of the gapped flat mode for several temperatures (0.05, 
0.1, 0.125, 0.15, 0.165, 0.175, 0.18, 0.185, 0.2 K) simulated using semi-classical molecular dynamics averaging over {\bf Q} from (0.1 0.1 0) to (0.9 0.9 0). (b-c) Temperature dependences fitted from (a) (see equation \ref*{fit_Xu}): (b) Temperature dependence of the intensities $I_0$ and $I'_0$ of the flat spin ice band and of the elastic contributions respectively. (c) Temperature dependence of the energy $E_0$ of the flat spin ice band. }
\end{figure*}

The main text of the present work compares the measured temperature dependence of $E_0$ and $I_0$ in our three samples with Monte Carlo calculations reported in Ref. \onlinecite{Xu20}. These calculations use effective exchange parameters from Ref. \onlinecite{Xu19}, which are detailed in Table 1 of the main text. They give a N\'eel temperature of 0.18 K. 

From these calculations, as illustrated in the Figure 3 of Ref. \onlinecite{Xu20}, two contributions are obtained: a spin ice elastic contribution which projects onto the ${\bf z}$ axis with a factor $\sin^2 \theta$, as well as spin waves features characteristic of the AIAO phase, with especially a flat spin ice band. To compare quantitatively these results with our data, those theoretical curves have been fitted to two modes, following:
\begin{equation}
I(E) = I_0~e^{-4 \log 2 (\frac{E-E_0)}{\delta_0})^2} + \frac{I'_0}{1+(\frac{E}{\delta'_0})^2}
\label{fit_Xu}
\end{equation}
The result of this fit is illustrated in Figures \ref*{App-Benton}(b) and \ref*{App-Benton}(c), which display respectively the intensity of the modes ($I_0$ and $I'_0$) and the position $E_0$ of the flat spin ice band. Interestingly, this energy remains finite even above the calculated critical temperature $T_{\rm N}=0.18$ K. 


\section{Correlations in magnetization measurements}
\begin{figure*}[h]
\includegraphics[width=\textwidth]{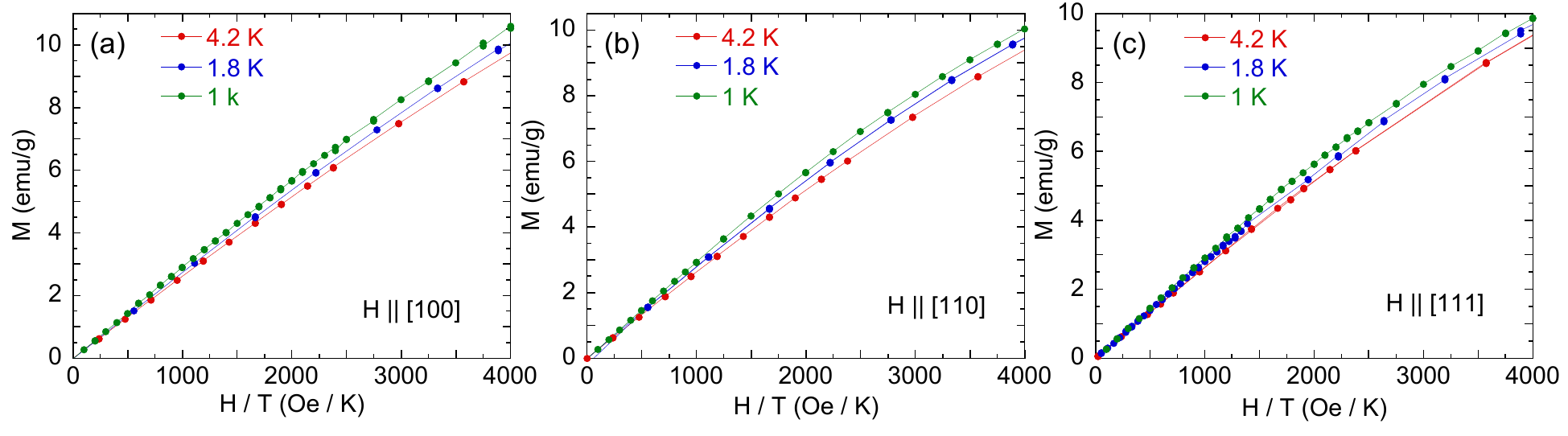}
\caption{\label{fig_MHT} Magnetization $M$ vs $H/T$ measured for Sample \#1 with the field applied along:  (a) [100], (b) [110] and (c) [111] at 1, 1.8 and 4.2 K.  }
\end{figure*}
In a paramagnet, isothermal magnetization curves scale as a function of the variable $H/T$. The deviations to this scaling give insight into the nature of the correlations that develop in the system. Upon cooling, if the magnetization curve increases faster (slower) than the higher temperature curve, it is the signature of the development of ferromagnetic-like (antiferromagnetic-like) correlations. 

We have plotted the magnetization as a function of $H/T$ for \ndzr, measured in Sample \#1.  As shown in Figure~\ref*{fig_MHT}, the $M(H/T)$ curves 
rise above the 4.2 K curves upon cooling down to 1 K, which indicates the 
development of ferromagnetic correlations, consistent with the elastic spin ice picture inferred from our neutron scattering measurements.  

At 500 mK, the curves lie between the 1 and 4.2 K curves (not shown on the figure for clarity), showing the development of antiferromagnetic correlations compared to 1 K, but the persistence of global ferromagnetic correlations. These antiferromagnetic correlations will end in the ``all in -- all out" ordering at about 300 mK. 


\end{document}